\newif\ifmulticol	\multicoltrue
\newif\ifshowgit	\showgitfalse		
\newif\ifgitlocal	\gitlocalfalse		
\newif\ifbiblatex	\biblatexfalse		
\newif\ifbibnum		\bibnumtrue 		
\newif\iflineno		\linenofalse
\newif\iftoc		\tocfalse

\newif\iflucida		\lucidafalse
\newif\ifcm			\cmfalse
\newif\iflibertine	\libertinefalse
\newif\ifcharter	\chartertrue

\newcommand*{\secnumdepth}{0}			
\newcommand*{\tocdepth}{2}				
\newcommand*{\reftitle}{References}		
\newcommand*{\mytitleskip}{-0.2in}		

\multicolfalse

\multicoltrue\showgittrue\gitlocaltrue\biblatexfalse\bibnumtrue

\newcommand*{\mydocfontsize}{\ifcharter11pt\else\iflibertine11pt\else10pt\fi\fi}
\newcommand*{\setcol}{\ifmulticol twocolumn\else onecolumn\fi}

\documentclass[\mydocfontsize,\setcol]{article}

\newcommand*{\pdfauthor}{Steven A.~Frank}
\newcommand*{\pdftitle}{Invariance in ecological pattern}

\input invariance.sty

\newcommand{\tr}{T}
\newcommand{\trz}{\tr_z}
\newcommand{\trr}{\tr_r}
\newcommand{\nmo}{n^{-1}}

\newcommand*{\rt}{\tilde{r}}
\newcommand*{\at}{\tilde{a}}

\newcommand*{\Ga}{\alpha}
\newcommand*{\Gb}{\beta}
\newcommand*{\Gd}{\delta}
\newcommand*{\GD}{\Delta}
\newcommand*{\Ge}{\epsilon}
\newcommand*{\Gg}{\gamma}

\newcommand*{\Gl}{\lambda}

\newcommand*{\Gm}{\mu}

\newcommand*{\Gs}{\sigma}
\newcommand*{\Gt}{\tau}
\newcommand*{\Gth}{\theta}

\DeclarePairedDelimiter\abs{\lvert}{\rvert}
\DeclarePairedDelimiter\norm{\lVert}{\rVert}
\DeclarePairedDelimiter\angb{\langle}{\rangle}
\DeclarePairedDelimiter\lrb{\lbrack}{\rbrack}
\DeclarePairedDelimiter\lr{\lparen}{\rparen}

\makeatletter
\let\oldabs\abs \def\abs{\@ifstar{\oldabs}{\oldabs*}}
\let\oldnorm\norm \def\norm{\@ifstar{\oldnorm}{\oldnorm*}}
\let\oldangb\angb \def\angb{\@ifstar{\oldangb}{\oldangb*}}
\let\oldlrb\lrb \def\lrb{\@ifstar{\oldlrb}{\oldlrb*}}
\let\oldlr\lr \def\lr{\@ifstar{\oldlr}{\oldlr*}}
\makeatother

\newcommand*{\dd}{\textrm{d}}

\newcommand*{\Eq}[1]{eqn~\ref{eq:#1}}

\newcommand*{\ovr}[2]{{{#1}\over{#2}}}
\newcommand*{\dovr}[2]{\ovr{\dd #1}{\dd #2}}

\newcommand*{\Figure}[1]{Figure~\ref{fig:#1}}
\newcommand*{\Fig}[1]{Fig.~\ref{fig:#1}}

\newcount\BoxNum \BoxNum 1\relax
\makeatletter
\newcommand*{\boxlabel}[1]{%
  \protected@write \@auxout {}{\string \newlabel {box:#1}{{\the\BoxNum}}{}}%
  \advance\BoxNum 1\relax}
\makeatother

\newcommand*{\boldrule}{\hrule height 1.2pt}

\newcommand*{\noterulenotec}[1]{\bigskip\boldrule\nobreak\medskip\nobreak%
	\centerline{\bf{\noindent #1}}\nobreak%
	\medskip\nobreak\boldrule\medskip}

\newcommand*{\mytitle}{\pdftitle}

\newcommand*{\myauthor}{Steven A.\ Frank}
\newcommand*{\myauthorb}{Jordi Bascompte}
\newcommand*{\myaddress}{Department of Ecology and Evolutionary Biology, University of California, Irvine, CA 92697--2525, USA}
\newcommand*{\myaddressb}{Department of Evolutionary Biology and Environmental Studies, University of Zurich, 8057 Zurich, Switzerland}

\setabstract{-0.07}{\iftoc-2.0\else1.22\fi}{%
The abundance of different species in a community often follows the log series distribution. Other ecological patterns also have simple forms. Why does the complexity and variability of ecological systems reduce to such simplicity? Common answers include maximum entropy, neutrality, and convergent outcome from different underlying biological processes. This article proposes a more general answer based on the concept of invariance, the property by which a pattern remains the same after transformation. Invariance has a long tradition in physics. For example, general relativity emphasizes the need for the equations describing the laws of physics to have the same form in all frames of reference. By bringing this unifying invariance approach into ecology, we show that the log series pattern dominates when the consequences of processes acting on abundance are invariant to the addition or multiplication of abundance by a constant. The lognormal pattern dominates when the processes acting on net species growth rate obey rotational invariance (symmetry) with respect to the summing up of the individual component processes. Recognizing how these invariances connect pattern to process leads to a synthesis of previous approaches. First, invariance provides a simpler and more fundamental maximum entropy derivation of the log series distribution. Second, invariance provides a simple derivation of the key result from neutral theory: the log series at the metacommunity scale and a clearer form of the skewed lognormal at the local community scale. The invariance expressions are easy to understand because they uniquely describe the basic underlying components that shape pattern.  
}

\begin{document}

\mymaketitle

\iftoc\mytoc{-24pt}{\newpage}\fi

{\em\noindent “It was Einstein who radically changed the way people thought about nature, moving away from the mechanical viewpoint of the nineteenth century toward the elegant contemplation of the underlying symmetry [invariance] principles of the laws of physics in the twentieth century” (ref.~\citenum{lederman04symmetry}, p.~153).}

\section{Introduction}

Ecologists have been interested in species abundance distributions (SADs) since the classic papers by Fisher\autocite{fisher43the-relation} and Preston\autocite{preston48the-commonness}. Two major patterns have been identified depending on the size of the community.  In a large community, abundances often follow the log series distribution \autocite{baldridge16an-extensive}. Specifically,  the probability that a species has a population size of $n$ individuals follows $p^n/n$. Communities differ only in their average population size, described by the parameter, $p$. At smaller spatial scales, the species abundance pattern often follows a skewed lognormal (a random variable is lognormally distributed when its logarithm is normally distributed) \autocite{hubbell01the-unified,rosindell11the-unified}. 

It is intriguing that the species abundance distribution follows these simple patterns irrespective of the particular group (birds, insects, mammals) and region considered. Other ecological patterns also follow simple probability distributions \autocite{brown95macroecology,gaston00pattern,harte11maximum}. Those patterns have attracted a lot of attention. Why does the variability and complexity of biology reduce to such a small range of simple distributions? How can we understand the relations between complex processes and simple patterns?

Approaches such as Harte's maximum entropy formalism \autocite{harte11maximum} and Hubbell's neutral theory \autocite{hubbell01the-unified} have attempted to explain the generality of the log series and skewed lognormal patterns in species abundance distributions. Maximum entropy describes probability distributions that are maximally random subject to satisfying certain constraints \autocite{jaynes57informationII,jaynes57information,jaynes03probability}. This approach has a long tradition in physics, both in statistical mechanics and information theory. An early maximum entropy approach in ecology derived the biomass pattern of populations\autocite{wagensberg87-the-extended,wagensberg88statistical,sherwin19the-introduction}. 

Neutral theory derives probability distributions by assuming that all individuals are equivalent \autocite{alonso08the-implicit}. Variation arises by random processes acting on the mechanistically identical individuals. Put another way, the mechanistic processes are ``neutral'' apart from random processes. Both maximum entropy and neutral theory have been shown to provide a good fit to the empirical patterns of species abundance distributions. In this article, we subsume these two different ways of understanding the log series and skewed lognormal patterns with a more general perspective based on the concept of invariance \autocite{frank16the-invariances}.  

\begin{figure*}[t]
\centering
\includegraphics[width=4in]{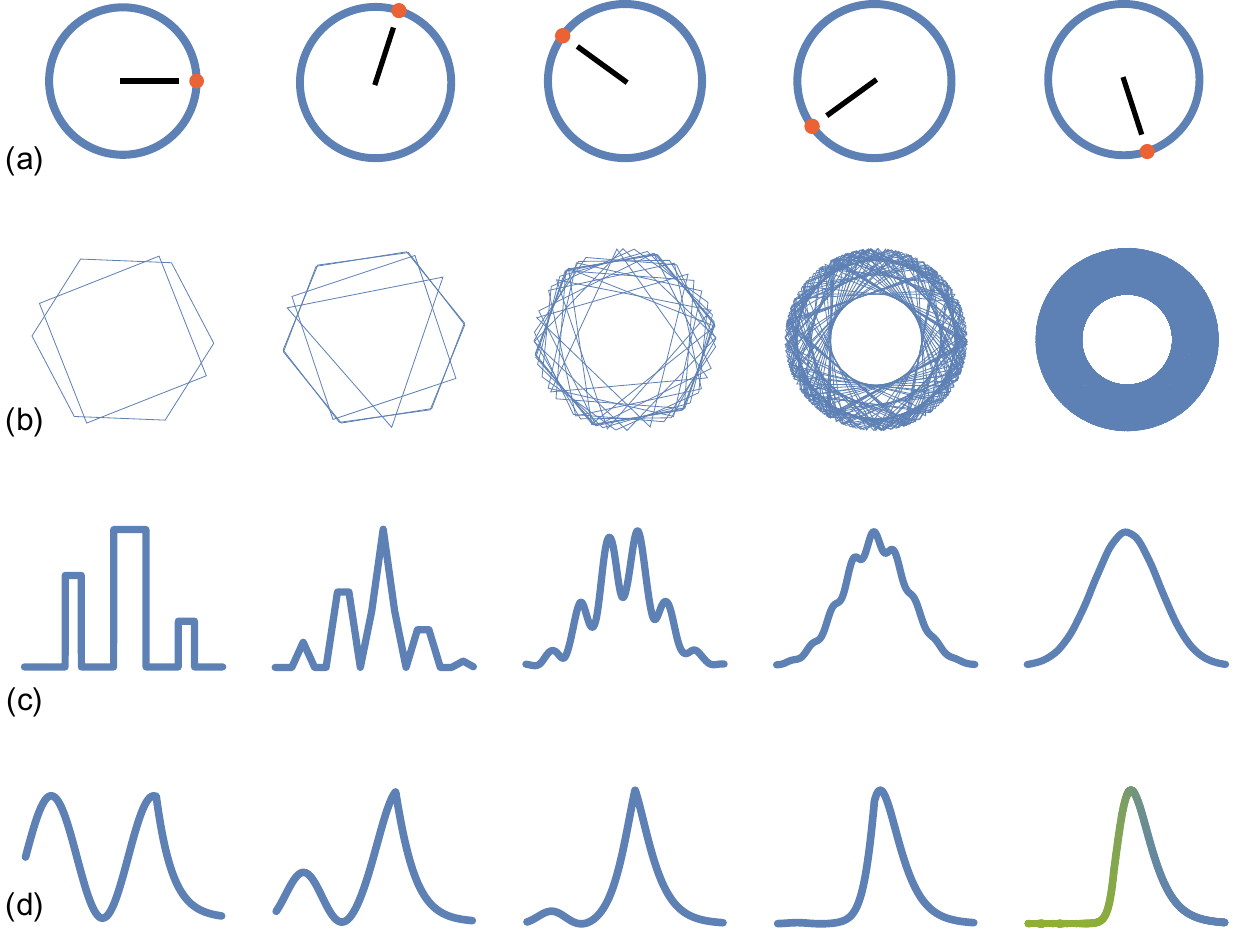}
\caption{Rotational and asymptotic invariance. (a) Transforming a circle by rotation leaves the circle unchanged (invariant), with an invariant radial distance at all points along the circumference. (b) Rotating regular polygons changes pattern. However, as more rotated polygons are added, the form converges asymptotically to a rotationally invariant circle, in which adding another rotated polygon does not change the pattern. Many common patterns of nature are asymptotically invariant. In this case, aggregation causes loss of all information except invariant radial distance. (c) The normal distribution is asymptotically invariant. The left curve describes an arbitrary probability pattern. The second curve expresses the sum of two randomly chosen values from the first curve. The height is the relative probability of the summed values. The third, fourth, and fifth curves express the sum of 4, 8, and 16 randomly chosen values from the first curve. Each curve width is shrunk to match the first curve. In this case, aggregation smooths the curve, causing loss of all information except the average squared distance from the center (the variance), which is equivalent to the average squared radial distance of rotationally invariant circles. (d) Extreme value distributions are asymptotically invariant. The left curve is an arbitrarily chosen probability pattern. The second curve expresses the probability of the largest value in a sample of two randomly chosen values from the first curve. The third, fourth, and fifth curves show the probability of the largest value of 4, 8, and 16 randomly chosen values. The asymptotically invariant curve on the right expresses exponential scaling at small values and linear scaling at large values, labeled in green and blue. Commonly observed probability distributions often express simple combinations of linear, logarithmic, and exponential scaling. Panels (a-c) modified from ref.~\citenum{frank16the-invariances}.} 
\label{fig:asymptotic}
\end{figure*}

\begin{figure*}[t]
\centering
\includegraphics[width=4in]{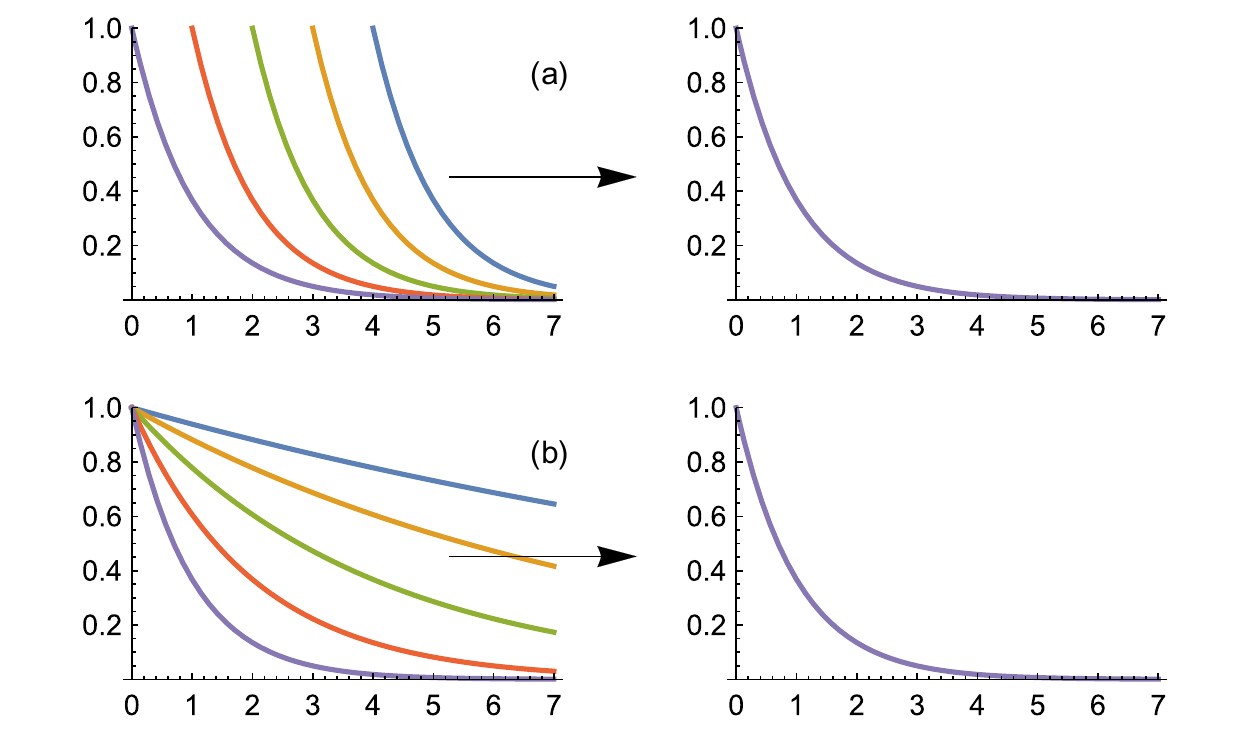}
\caption{Shift and stretch invariance of the exponential distribution. (a) The left panel shows $e^{-(x+a)}$ for $a=0,-1\dots,-4$. Decreasing values of $a$ shift the curve to the right, which is equivalent to shifting the $x$ axis by resetting the zero point. For probability patterns, the total probability must be normalized to one, which means that all curves must have the same area under the curve for values of $x$ between $0$ and $\infty$. To normalize the curves, the right panel plots $k_ae^{-(x+a)}$ with $k_a=e^a$. Thus, all curves become $e^{-x}$ invariantly with respect to different shift values, $a$. (b) The left panel shows $e^{-bx}$ for $b=2^0,2^{-1}\dots,2^{-4}$. Decreasing values of $b$ stretch the $x$ axis by a factor of $2$ for each halving of $b$. To normalize the average value of each probability curve to be the same, the right panel shows $e^{-\Gl_b bx}$ for $\Gl_b=1/b$. Thus, all curves become $e^{-x}$ invariantly with respect to different stretch values, $b$.}
\label{fig:affine}
\end{figure*}

Invariance can be defined as the property by which a system remains unchanged under some transformation. For example, a circle is the same (invariant) before and after rotation (\Fig{asymptotic}a). In ecology, pattern often depends on the ways in which form remains invariant to changes in measurement. Some patterns retain the same form after uniformly stretching or shrinking the scale of measurement (\Fig{affine}b). Measures of length provide a common example of stretch invariance. One can measure lengths equivalently in millimeters or centimeters without loss of information. As we will see, that kind of invariance often determines the form of observed pattern. 

To give another example, consider the common and widely familiar pattern of the normal distribution. By the central limit theorem, when independent random variables are added, their properly normalized sum tends toward a normal distribution, even when the component variables themselves are not normally distributed. The central limit theorem and the normal distribution are often considered as unique aspects of pattern that stand apart from other commonly observed patterns.

The invariance perspective that we promote shows how the normal distribution is in fact a specific example of a wider framework in which to understand the commonly observed patterns of nature. In particular, the normal distribution arises from the rotational invariance of the circle \autocite{frank16common}. For two variables, $x$ and $y$, with a given squared length, $x^2+y^2=r^2$, all combinations of the variables with the same radius, $r$, lie along the circumference of a circle (\Fig{asymptotic}a). When each combination is equally likely, the rotationally invariant radius is sufficient to describe the probability pattern.

It is this rotational invariance that gives the particular mathematical form of the normal distribution, in which the average squared radius sets the variance of the distribution. By this perspective, the mathematical forms of all commonly observed distributional patterns express their unique invariances\autocite{frank16common}.

The perspective of invariance was the basis for most of the great conceptual advances of physics in the twentieth century\autocite{lederman04symmetry}. For example, Gell-Mann's pioneering theoretical work on the fundamental particles of nature derived from invariance (symmetry) properties that unified understanding of known particles and predicted new particles such as quarks, which were subsequently observed. By contrast, general aspects of invariance have not been used consistently as the fundamental basis for understanding patterns in ecology. One exception concerns scale invariance, which is often discussed in ecology \autocite{sole99criticality,jordano03invariant,stouffer05quantitative}. But scale invariance is typically limited to special kinds of patterns rather than forming a unified approach to diverse patterns.

The point of this paper is that invariance is the most general way in which to understand commonly observed patterns. Species abundance distributions provide an excellent illustration. For species abundances, we show that maximum entropy and neutral models can succeed in certain cases because they derive from invariance principles. However, maximum entropy and neutrality are often difficult to interpret because they hide their underlying basis in invariance. 

Our unifying invariance analysis clarifies why seemingly different conceptual approaches have the same consequences for pattern. Similarly, seemingly different biological processes may often lead to the same observed pattern, because those different processes share a common basis in invariance. That deeper understanding suggests a more insightful way to think about alternative mechanistic models. It also suggests the kinds of empirical tests that may differentiate between alternative causal processes.

This manuscript is organized as follows. First, we highlight new theoretical results for species abundance distributions. Second, we review how invariance defines probability patterns in a general way \autocite{frank11a-simple,frank14how-to-read,frank16common}. The log series distribution\autocite{frank19the-common} and the new gamma-lognormal distribution for species abundances follow directly from the universal invariance expression of probability patterns. Third, we show that maximum entropy and neutrality can easily be understood as special examples of invariance principles. Finally, we discuss the broad role of invariance in the analysis of ecological pattern. 

\section{New results}

This article develops two new theoretical results. We highlight those results before starting on the general overview of invariance and pattern.

First, we present a simple maximum entropy derivation of the log series pattern. We show that constraining the average abundance per species is sufficient when analyzing randomness and entropy on the proper demographic scale. 

The simplicity of our maximum entropy derivation contrasts with Harte's more complicated maximum entropy model \autocite{harte08maximum,harte11maximum}. Harte had to assume an additional unnecessary constraint on energy usage. He required that unnecessary constraint because he evaluated randomness on the scale of measured abundances rather than on the scale of demographic process. This will be made explicit below. 

We use this new result to demonstrate that maximum entropy is the outcome of deeper underlying principles of invariance and pattern. By working at the deeper level of invariance, one obtains a simpler and more powerful understanding of pattern.

The second new result shows that Hubbell's \autocite{hubbell01the-unified} neutral model is the simple expression of three basic invariances. Hubbell's full range of log series and skewed lognormal (zero sum multinomial) results follows immediately from those three underlying invariances. 

The three invariances correspond to a maximum entropy model that constrains the average abundance of species and the average and variance of the demographic processes influencing abundance. The three invariances lead to a simple gamma-lognormal distribution that matches the neutral theory pattern for species abundances. The gamma-lognormal is a product of the standard gamma and lognormal distributions.

\section{Invariance}

This section reviews how invariance considerations lead to the log series distribution\autocite{frank19the-common}. We delay discussion of the gamma-lognormal until the later section on Hubbell's neutral model. 

\subsection{Canonical form of probability distributions}

We can rewrite almost any probability distribution as
\begin{equation}\label{eq:Tz}
  q_z=ke^{-\Gl\trz},
\end{equation}
in which $\tr(z)\equiv\trz$ is a function of the variable, $z$, and $k$ and $\lambda$ are constants. For example, Student's t-distribution, usually written as 
\begin{equation*}
  q_z=k\lr{1+z^2/\nu}^{-\lr{\nu+1}/2}
\end{equation*}
can be written in the form of \Eq{Tz} with $\Gl=\lr{\nu+1}/2$ and $\trz=\log\lr{1+z^2/\nu}$.

The probability pattern, $q_z$, is invariant to a constant shift, $\trz\mapsto a+\trz$, because we can write the transformed probability pattern in \Eq{Tz} as
\begin{equation*}
  q_z=k_ae^{-\Gl\lr{a+\trz}}=ke^{-\Gl\trz},
\end{equation*}
with $k=k_ae^{-\Gl a}$ (\Fig{affine}a). We express $k$ in this way because $k$ adjusts to satisfy the constraint that the total probability be one. In other words, conserved total probability implies that the probability pattern is shift invariant with respect to $\trz$ (see ref.~\citenum{frank16common}).

Now consider the consequences if the average of some value over the distribution $q_z$ is conserved. For example, the average of $z$ is the mean, $\Gm$, and the average of $\lr{z-\Gm}^2$ is the variance. A constraint causes the probability pattern to be invariant to a multiplicative stretching (or shrinking), $\trz\mapsto b\trz$, because
\begin{equation*}
  q_z=ke^{-\Gl_b b\trz}=ke^{-\Gl\trz},
\end{equation*}
with $\Gl=\Gl_b b$ (\Fig{affine}b). We specify $\Gl$ in this way because $\Gl$ adjusts to satisfy the constraint of conserved average value. Thus, invariant average value implies that the probability pattern is stretch invariant with respect to $\trz$.

Conserved total probability and conserved average value cause the probability pattern to be invariant to an affine transformation of the $\trz$ scale, $\trz\mapsto a+b\trz$, in which ``affine'' means both shift and stretch.

The affine invariance of probability patterns with respect to $\trz$ induces significant structure on the form of $\trz$ and the associated form of probability patterns. Understanding that structure provides insight into probability patterns and the processes that generate them  \autocite{frank11a-simple,frank14how-to-read,frank16common}.

In particular, Frank \& Smith \autocite{frank11a-simple} showed that the invariance of probability patterns to affine transformation, $\trz\mapsto a+b\trz$, implies that $\trz$ satisfies the differential equation
\begin{equation*}
  \dovr{\trz}{w}=\Ga+\Gb\trz,
\end{equation*}
in which $w(z)$ is a function of the variable $z$. The solution of this differential equation expresses the scaling of probability patterns in the generic form
\begin{equation}\label{eq:wmetric}
  \trz\stackrel{}{=}\frac{1}{\Gb}\lr{e^{\Gb w}-1},
\end{equation}
in which, because of the affine invariance of $\trz$, we have added and multiplied by constants to obtain a convenient form, with $\trz\rightarrow w$ as $\Gb\rightarrow0$. 

By writing $\trz$ in this way, $w$ expresses a purely shift-invariant aspect of the fundamental affine-invariant scale, because the shift transformation $w\mapsto a+w$ multiplies $\trz$ by a constant, and probability pattern is invariant to constant multiplication of $\trz$. Thus, \Eq{wmetric} dissects the anatomy of a probability pattern (\Eq{Tz}) into its component invariances.

With this expression for $\trz$, we may write probability patterns generically as
\begin{equation}\label{eq:canonical}
  q_z=ke^{-\Gl \lr{e^{\Gb w}-1}/\Gb}.
\end{equation}
This form has the advantage that $w(z)$ expresses the shift-invariant structure of a probability pattern. Most of the commonly observed probability patterns have a simple form for $w$ \autocite{frank14how-to-read,frank16invariant}. That simplicity of the shift-invariant scale suggests that focus on $w$ provides insight into common patterns.

\subsection{Proportional processes and species abundances}

To understand the log series, we must consider the relation $n=e^r$ between the observed pattern of abundances, $n$, and the processes, $r$. Here, $r$ represents the total of all proportional processes acting on abundance \autocite{frank19the-common}. 

A proportional process simply means that the number of individuals or entities affected by the process increases in proportion to the number currently present, $n$. Demographic processes, such as birth and death, act proportionally. 

The sum of all of the proportional processes on abundance over some period of time is
\begin{equation*}
  r = \int_0^\Gt m(t)\dd t.
\end{equation*}
Here, $m(t)$ is a proportional process acting at time $t$ to change abundance. Birth and death typically occur as proportional processes. The value of $r=\log n$ is the total of the $m$ values over the total time, $\Gt$. For simplicity, we assume $n_0=1$.

The log series follows as a special case of the generic probability pattern in \Eq{canonical}. To analyze abundance, focus on the process scale by letting the variable of interest be $z\equiv r$, with the key shift-invariant scale as simply the process variable itself, $w(r)=r$. Then \Eq{canonical} becomes
\begin{equation}\label{eq:canonicalR}
  q_r\dd r=ke^{-\Gl \lr{e^{\Gb r}-1}/\Gb}\,\dd r,
\end{equation}
in which $q_r\dd r$ is the probability of a process value, $r$, in the interval $r+\dd r$. 

Using $w(r)=r$ sets the the shift-invariant scale as the variable itself. Substituting this simplest form for the shift-invariant scale into the canonical equation for common probability patterns in \Eq{canonical} yields the simplest generic expression of probability pattern as \Eq{canonicalR}. 

We can generalize the relation between abundance and process, $n=e^r$, by writing $n^\Gb=e^{\Gb r}$, which uses an additional parameter $\Gb$ to allow comparison with the canonical form of probability distributions in the previous subsection. When we focus on standard models of species abundances, we use $\Gb=1$.

We can change from the process scale, $r$, to the abundance scale, $n$, by noting that $\Gb\log n=\Gb r$, and so, for any $\Gb$, we have $r=\log n$. Thus, we can use the substitutions $r\mapsto\log n$ and $\dd r\mapsto \nmo\dd n$ in \Eq{canonicalR}, yielding the identical probability pattern expressed on the abundance scale
\begin{equation}\label{eq:canonicalN}
  q_n\dd n=k\nmo e^{-\Gl \lr{n^\Gb-1}/\Gb}\,\dd n.
\end{equation}
The value of $k$ always adjusts to satisfy the constraint of invariant total probability, and the value of $\Gl$ always adjusts to satisfy the constraint of invariant average value. 

For proportional processes and species abundances, $\Gb=1$, as noted above. For that value of $\Gb$, we obtain the log series distribution \autocite{frank19the-common}
\begin{equation}\label{eq:logseriesN}
  q_n=k\nmo e^{-\Gl n},
\end{equation}
replacing $n-1$ by $n$ in the exponential term which, because of affine invariance, describe the same probability pattern. The log series is often written with $e^{-\Gl}=p$, and thus $q_n=kp^n/n$. One typically observes discrete values $n=1,2,\dots$. See the Appendix for the general relation between discrete and continuous distributions. The continuous analysis here is sufficient to understand pattern. 

We can also write the log series on the process scale, $r$, from \Eq{canonicalR}, as \autocite{frank19the-common}
\begin{equation}\label{eq:logseriesR2}
  q_r=ke^{-\Gl e^r}.
\end{equation}
This form shows that the log series is the simplest expression of generic probability patterns in \Eq{canonical}. The log series arises from $\Gb=1$, associated with $n=e^r$, and from the base shift-invariant scale as $w\equiv r$ for proportional processes, $r$. 
 
\subsection{Invariances of the log series}

This subsection summarizes a few technical points about invariance. These technical points provide background for our simpler and more general derivation in the following section of maximum entropy models for species abundances. Those prior models focused only on abundances, $n$, without considering the underlying process scale, $r$.

We begin with invariance on the process scale, $r$. On that scale, the log series in \Eq{logseriesR2} is the pure expression of additive shift invariance to $r$ and lack of multiplicative stretch invariance to $r$. For example, note in \Eq{logseriesR2} that an additive change, $r\mapsto r+a$, is compensated by a change in $\Gl$ to maintain the overall invariance, whereas a multiplicative change, $r\mapsto br$, cannot be compensated by a change in one of the constants. For example, if $r$ is net reproductive rate, then an improvement in the environment that adds a constant to everyone's reproductive rate does not alter the log series pattern. By contrast, multiplying reproductive rates by a constant does alter pattern.

To understand the parameter, $\Gb$, from \Eq{wmetric}, consider that
\begin{equation*}
  \tr=\frac{1}{\Gb}\lr{e^{\Gb r}-1}=\frac{1}{\Gb}\lr{n^\Gb-1},
\end{equation*}
in which $\Gb$ is the relative curvature of the measurement scale for abundance, $n$, with respect to the scale for process, $r$. The relative curvature is $\Gb=\tr''/\tr'$, with the primes denoting differentiation with respect to $r$. 

For the log series, the curvature of $\Gb=1$ describes the amount of bending of the abundance scale, $n=e^r$, with respect to multiplying the process scale, $r$, by a constant---the departure from stretch invariance. 

The simple invariances with respect to process, $r$, become distorted and more difficult to interpret when we focus only on the observed scale for abundance, $n$, associated with the log series in \Eq{logseriesN}. In that form of the distribution, the canonical scale is
\begin{equation}\label{eq:Tn}
  \tr=\frac{1}{\Gl}\log n + n.
\end{equation}
In this expression, purely in terms of abundances, the logarithmic term dominates when $n$ is small, and the linear term dominates when $n$ is large. Thus, the scale changes from stretch but not shift invariant at small magnitudes to both shift and stretch invariant at large magnitudes \autocite{frank19the-common}. 

Without the simple insight provided by the process scale, $r$, we are left with a complicated and nonintuitive pattern that is separated from its simple cause. That difficulty has led to unnecessary complications in maximum entropy theories of pattern.

\section{Maximum entropy}

Maximum entropy describes probability distributions that are maximally random subject to satisfying certain constraints \autocite{jaynes57information,jaynes57informationII,jaynes03probability}. In \Eq{Tz}, with the generic description for distributions as
\begin{equation*}
  q_z\dd z=ke^{-\Gl \trz}\dd z,
\end{equation*}
maximum entropy interprets this form as the expression of maximum randomness with respect to the scale $z$, subject to the constraint that the average of $\trz$ is fixed \autocite{frank14how-to-read}.

This section begins with a maximum entropy derivation for the log series based on our separation between the scales of process, $r$, and observed abundance, $n$. 

We then discuss Harte's \autocite{harte08maximum,harte11maximum} alternative maximum entropy derivation of the log series. Harte's derivation emphasizes mechanistic aspects of energy constraints rather than our emphasis on the different scales of process and abundance. 

\subsection{Constraint of average abundance on process scale}

The log series in \Eq{logseriesR2} is
\begin{equation*}
  q_r\dd r=ke^{-\Gl e^r}\dd r.
\end{equation*}
Here, $\tr=e^r=n$. This distribution expresses maximum entropy with respect to the process scale, $r$. The constraint is the ecological limitation on average abundance
\begin{equation}\label{eq:nrconstraint}
  \angb{\tr}_r=\angb{e^r}_r=\angb{n}_r,
\end{equation}
in which $\angb{\cdot}_r$ denotes average value with respect to the process scale, $r$. 

In this case, process values, $r$, are maximally random, subject to the ecological constraint that limits abundance, $n$. Thus, maximizing entropy with respect to the process scale, $r$, subject to a constraint on the observed pattern scale, $n$, leads immediately to the log series.

Relating the process scale, $r$, to the scale of ecological constraint, $n$, often makes sense. Typically, environmental perturbations associate with changes in demographic variables, such as birth and death rates. Such demographic factors typically act proportionally on populations, consistent with our interpretation of $r$ as the aggregate of proportionally acting processes. The perturbations, acting on demographic variables, associate the process scale with the scale of randomness. 

In contrast with the process scale of perturbation and randomness for the demographic variables, the scale of constraint naturally arises with respect to a limit on the number of individuals, $n$. Thus, randomness happens on the $r$ scale and constraint happens on the $n$ scale.

It is, of course, possible to formulate alternative models in which randomness and constraint happen on scales that differ from our interpretation. Different formulations are not intrinsically correct or incorrect. Instead, they express different assumptions about the relations between process, randomness, and invariance. The next section considers an alternative formulation.

\subsection{Harte's joint constraints of abundance and energy}

Harte developed comprehensive maximum entropy models of ecological pattern. He tested those theories with the available data. His work synthesizes many aspects of ecological pattern \autocite{harte11maximum}.

For species abundances, Harte \autocite{harte08maximum,harte11maximum} analyzed maximum randomness with respect to the scale of abundance values, $n$. Maximum entropy derivations commonly evaluate randomness on the same scale as the observations. In this case, with observations for the probabilities of abundances, $p_n$, entropy on the same scale is the sum or integral of $-p_n\log p_n$.

However, there is no a priori reason to suppose that the scale of observation is the same as the scale of randomness. The fact that observation, randomness, and process may occur on different scales often makes maximum entropy models difficult to develop and difficult to interpret. For example, we may observe the probabilities of abundances, $p_n$, but randomness may be maximized on the scale of process, as the sum or integral of $-p_r\log p_r$.

In the final part of this section, we argue that invariance provides a truer path to the natural scale of analysis and to the mechanistic processes that generate pattern than does maximum entropy. Before comparing invariance and maximum entropy, it is useful to sketch the details of Harte's maximum entropy model for species abundances.

The simplest maximum entropy model analyzes entropy with respect to abundance, $n$, subject to a constraint on the average abundance, $\angb{n}$. That analysis yields an exponential distribution
\begin{equation*}
  q_n\dd n=ke^{-\Gl n}\dd n.
\end{equation*}
The exponential pattern differs significantly from the observed log series pattern. Thus, maximizing entropy with respect to the scale of abundance, $n$, and constraining the average abundance is not sufficient.

From our invariance perspective, it is natural to think of the scale of randomness in terms of $\dd r$, the scale of proportional processes, rather than in terms of $\dd n$, the scale of abundance. Maximizing randomness with respect to $\dd r$ leads directly to the log series, as shown in the previous section.

Harte did not consider the distinction between the exponential and log series patterns with respect to the scale of randomness. Instead, to go from the default exponential pattern of maximum entropy to the log series, his maximum entropy analysis required additional assumptions. He proceeded in the following way.

Suppose that the total quantity of some variable, $\Ge$, is constrained to be constant over all individuals of all species. The average value per individual is $\angb{\Ge}$. It does not matter what the variable $\Ge$ is. All that matters is that the constraint exists. Harte assumed that $\Ge$ is energy, but that assumption is unnecessary with regard to the species abundance distribution.

The value $\Ge$ is distributed over individuals independently of their species identity. Thus, the variable $\Gd|n=n\Ge$ is the total value in a species with $n$ individuals, with average value $\angb{\Gd|n}=n\angb{\Ge}$. 

The joint distribution of $n$ and $\Gd$ is 
\begin{equation*}
  q_{n,\Gd}=q_nq_{\Gd|n}.
\end{equation*}
The explicit form of this joint distribution can be obtained by maximizing entropy subject to the constraints on the average abundance per species, $\angb{n}$, and the average total value in a species with $n$ individuals, $\angb{\Gd|n}$, yielding
\begin{equation*}
  q_{n,\Gd}=ke^{-\Gl n}e^{-\Gl'\Gd}.
\end{equation*}
We obtain the form presented by Harte \autocite{harte08maximum} using the equivalence $\Gd= n\Ge$, yielding
\begin{equation*}
  q_{n,\Ge}=ke^{-\Gl n}e^{-\Gl'n\Ge}.
\end{equation*}
The species abundance distribution is obtained by 
\begin{equation*}
  q_n=\int q_{n,\Ge}\dd\Ge=ke^{-\Gl n}\int e^{-\Gl'n\Ge}\dd\Ge.
\end{equation*}
Noting that $\int e^{-\Gl'n\Ge}=1/\Gl'n$, and absorbing the constant $\Gl'$ into $k$, we obtain the log series for the species abundance distribution
\begin{equation*}
  q_n=k\nmo e^{-\Gl n}.
\end{equation*}

\subsection{Maximum entropy and invariance}

Harte's maximum entropy derivation of the log series assumes joint constraints of abundance, $n$, and some auxiliary variable, $\Ge$, which he labeled as energy. He evaluated entropy on the scales of $n$ and $\Ge$.

By contrast, our invariance derivation arises from a constraint on abundance plus evaluation of invariance or entropy on the scale $r=\log n$. On that scale, the log series arises in a simple and clear way. There is no need for constraint of a second auxiliary variable. 

Without an invariance argument, nothing compels us to analyze with respect to the $r$ scale. Harte, without focus on invariance, followed the most natural approach of using $n$ as the scale for maximization of randomness and for constraint. That approach required an auxiliary constraint on a second scale to arrive at the log series. 

Harte's approach was a major step in unifying the analysis of empirical pattern. But, in retrospect, his approach was unnecessarily complicated.

One might say that Harte's approach provided a richer theory because it led to predictions about both abundance and energy. However, the data on abundance patterns match very closely to the log series, whereas the data for different proxies of energy vary considerably \autocite{harte11maximum}. 

Our invariance approach strips away the unnecessary auxiliary variable. The invariance theory therefore provides a much simpler way to derive and to understand abundance patterns.

Maximum entropy can be thought of purely as a basic invariance method of analysis. Maximum entropy distributions have the form in \Eq{Tz} as
\begin{equation*}
  q_z\dd z=ke^{-\Gl \trz}\dd z,
\end{equation*}
in which $\trz$ is the affine-invariant scale that defines the probability pattern. Thus, the method of maximum entropy is simply a method for deriving the affine-invariant expression, $\trz$. In practice, maximum entropy has three limitations. 

First, maximum entropy is silent with respect to the proper choice for the scale on which entropy is maximized and the constraints that set the affine-invariant expression, $\trz$. By contrast, focus on invariance led us to the shift invariance of the process scale, $r$. That scale provided a much simpler analysis, in which $r$ is the incremental scale with respect to invariance and the measurement scale with respect to entropy. 

In other words, maximum entropy is a blind application of the most basic invariance principles, without any guidance about the proper scales for invariance, randomness, and constraint. By contrast, an explicit invariance approach takes advantage of the insight provided by the analysis of invariance.

Second, by focusing on invariance, we naturally obtain the full invariance (symmetry) group expression in \Eq{canonical} as the generic form of probability patterns
\begin{equation*}
    q_z=ke^{-\Gl \lr{e^{\Gb w}-1}/\Gb}.
\end{equation*}
That generic expression leads us to a generalization of the log series in \Eq{canonicalN} as \autocite{frank19the-common}
\begin{equation*}
  q_n\dd n=k\nmo e^{-\Gl \lr{n^\Gb-1}/\Gb}\,\dd n,
\end{equation*}
which is a two parameter distribution for abundances with respect to $\Gl$ and $\Gb$. The log series is a special case with $\Gb=1$.

Third, invariance leads to a deeper understanding of the relation between observed pattern and alternative mechanistic models of process. The following section provides an example.

\section{Neutrality}

Here, we analyze Hubbell's \autocite{hubbell01the-unified} neutral model of species abundances in the light of our invariance perspective. With that example in mind, we then discuss more generally how neutral models relate to invariance and maximum entropy. 

\subsection{Hubbell's neutral model}

The strong recent interest in Hubbell's neutral model follows from the match of the theory to the contrasting patterns of species abundance distributions (SADs) that have been observed at different spatial scales. In the theory, many local island-like communities are connected by migration into a broader metacommunity. Sufficiently large metacommunities follow the log series pattern of species abundances. Each local community follows a distribution that Hubbell called the zero-sum multinomial\autocite{chisholm10niche}, which is similar to a skewed lognormal. As noted by Rosindell et al.\autocite{rosindell11the-unified}, it is this flexibility of the classic neutral model to reconcile the log series and lognormal distributions that allows it to fit empirical data well \autocite{etienne11the-spatial}.

\begin{figure*}[t]
\centering
\includegraphics[width=4in]{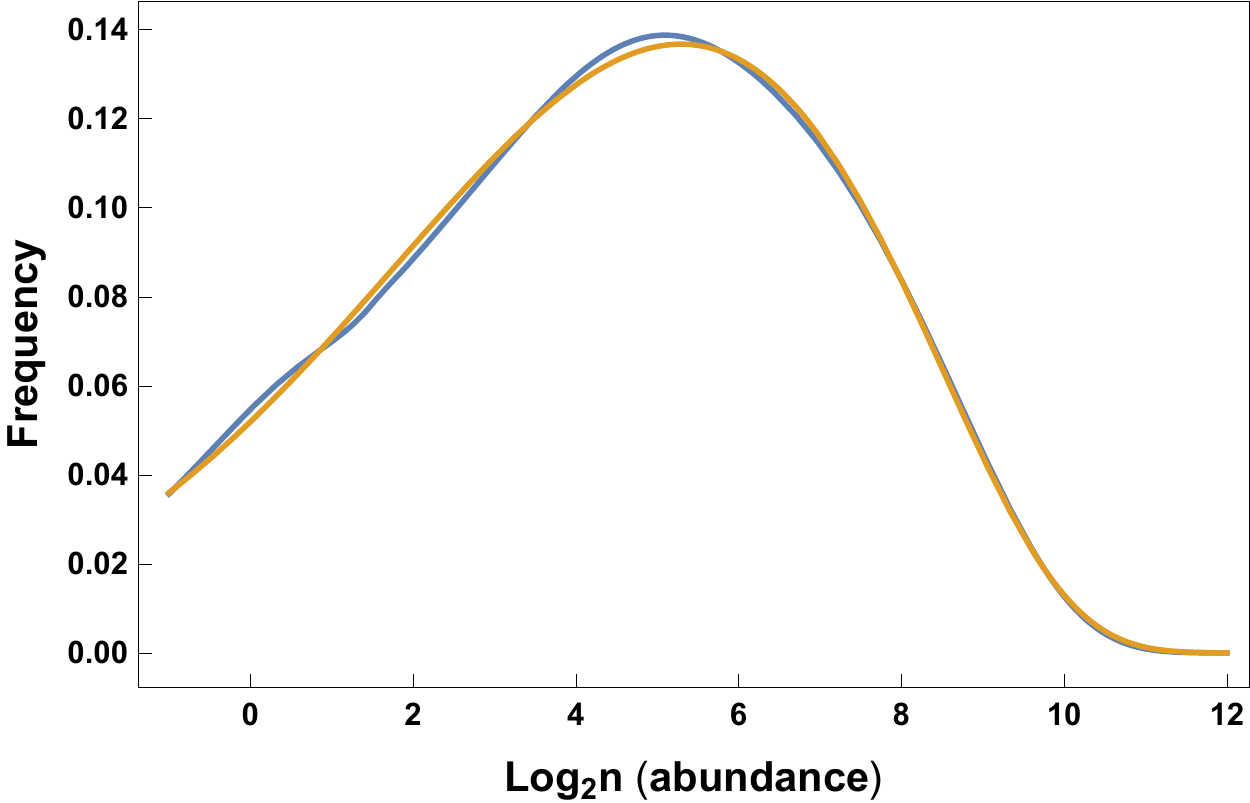}
\caption{Match of the gamma-lognormal pattern in gold to the neutral theory fit in blue for Panama tree species abundances. The neutral theory fit to the data comes from Chisholm and Pacala's \autocite{chisholm10niche} analysis in their Fig.~1. They used Hubbell's neutral theory model with parameters $J=21,060$, $m=0.075$, and $\Gth=52.1$ in their eqn~3, originally from Alonso and McKane \autocite{alonso04sampling}. The gamma-lognormal model in \Eq{gammaLNR} produces essentially the identical pattern with parameters $\Gl=0.00205$, $a=0.491$, and $\Ga=0.0559$. The abundance scale can be expressed equivalently on the process scale, $\log_2 n = r/\log 2$. See the Data accessibility statement for the calculations used to produce this plot.} 
\label{fig:pacalaFit}
\end{figure*}

\subsection{Invariance and the gamma-lognormal distribution}

Broad consensus suggests that species abundances closely follow the log series pattern at large spatial scales. Extensive data support that conclusion \autocite{baldridge16an-extensive}.

Observed pattern at small spatial scales differs from the log series. Consensus favors a skewed lognormal pattern. The data typically show an excess of rare species, causing a skew relative to the symmetry of the lognormal when plotted on a logarithmic scale. 

At small spatial scales, most recent analyses focus on data from a single long-term study of tree species in Panama \autocite{hubbell01the-unified,chisholm10niche}. Thus, some ambiguity remains about the form and consistency of the actual pattern at small scales.

The blue curve of \Fig{pacalaFit} shows Chisholm and Pacala's \autocite{chisholm10niche} fit of the neutral theory to the Panama tree data for species abundances at small spatial scales. The gold curve shows the close match to the neutral theory pattern by a simple probability distribution derived from the analysis of invariance.

To obtain the matching distribution derived by invariance, we begin with the canonical form for probability distributions in \Eq{canonical}. That canonical form expresses pattern in terms of the shift-invariant scale, $w$. Next, we need to find the specific form of the scale $w$ that relates this canonical form for probability distributions to the neutral theory. Because the neutral theory derives abundance, $n$, as an outcome of demographic processes, $r$, the fundamental shift-invariant scale for neutral theory is expressed in terms of the demographic process variable as
\begin{equation}\label{eq:wr}
  w=\log\lr{e^r - \frac{a}{\Gl}r+\frac{\Ga}{\Gl} r^2}.
\end{equation}
Below, we discuss why this is a natural shift-invariant scale for neutral theory. For now, we focus on the details of the mathematical expressions. Recall that $n=e^r$ relates measured abundances, $n$, to the demographic process scale, $r$. If we assume that $\Gb=1$ in \Eq{canonical} and use $w$ from \Eq{wr}, we obtain
\begin{equation}\label{eq:gammaLNR}
  q_r=ke^{-\Gl e^r + ar - \Ga r^2},
\end{equation}
with parameters $\Gl$, $a$, and $\Ga$. We can write this distribution equivalently on the $n$ scale for abundance as
\begin{equation}\label{eq:gammaLNN}
  q_n=kn^{\at-1}e^{-\Gl n}e^{-\Ga\lr{\log n - \Gm}^2}.
\end{equation}
In the second distribution, $\Gm=(a-\at)/2\Ga$. Thus, both distributions have the same three parametric degrees of freedom. 

The right-hand exponential term of \Eq{gammaLNN} is a lognormal distribution with parameters $\Gm$ and $\Gs^2=1/2\Ga$. The remaining terms are a gamma distribution with parameters $\at$ and $\Gl$. We call this product of the gamma and lognormal forms the gamma-lognormal distribution. 

\begin{figure*}[t]
\centering
\includegraphics[width=5.5in]{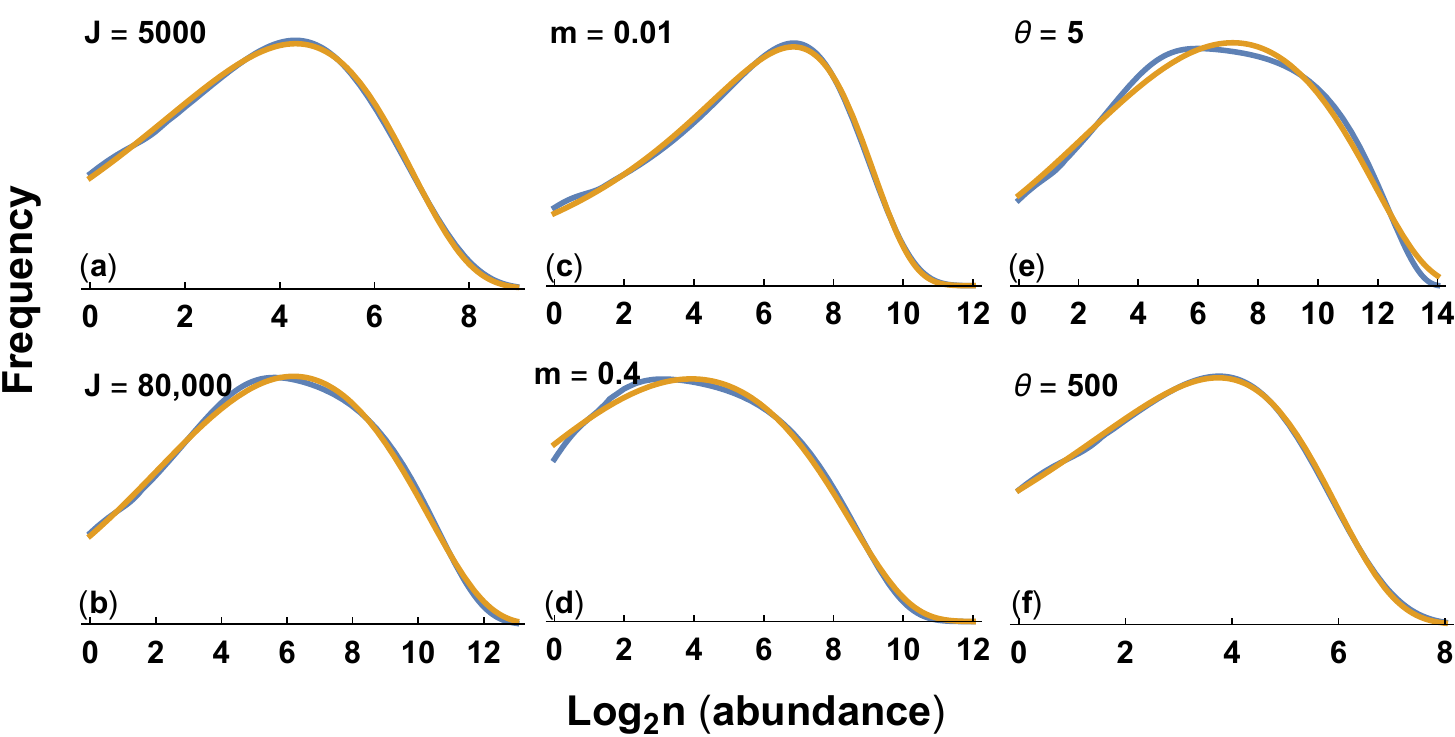}
\caption{Match of Hubbell's neutral theory to the gamma-lognormal distribution. The blue curve for the neutral theory and the gold curve for the gamma-lognormal are calculated as described in \Fig{pacalaFit}. The parameters for the neutral theory are the same as in \Fig{pacalaFit}, except as shown in each panel. I fit the parameters for the gamma-lognormal to each neutral theory curve, with values for each panel: (a) $\Gl=0.01115$, $a=0.4452$, and $\Ga=0.03660$; (b) $\Gl=0.0004209$, $a=0.4622$, and $\Ga=0.05014$; (c) $\Gl=0.002765$, $a=0.3182$, and $\Ga=0$; (d) $\Gl=0.001777$, $a=0.2217$, and $\Ga=0.03576$; (e) $\Gl=0.0001509$, $a=0.3851$, and $\Ga=0.03667$; (f) $\Gl=0.02519$, $a=0.3726$, and $\Ga=0.006900$. See the Data accessibility statement for the calculations used to produce these plots.} 
\label{fig:neutralFit}
\end{figure*}

\Figure{pacalaFit} showed that the gamma-lognormal distribution matches the neutral theory fit for the Panama tree data. \Fig{neutralFit} shows that the shape of the gamma-lognormal matches the shape of the neutral theory predictions for various mechanistic parameters of the neutral theory. 

In summary, the neutral theory distribution appears to be nearly identical to a gamma-lognormal distribution when compared over realistic parameter values. Both distributions have the same three parametric degrees of freedom. 

\subsection{Maximum entropy and the gamma-lognormal}

The constraints on pattern can be seen most clearly by rewriting \Eq{gammaLNR} as
\begin{equation}\label{eq:gammaLNRu}
  q_r=ke^{-\Gl\trr}=ke^{-\Gl e^r + \at r - \Ga \rt^2},
\end{equation}
in which $\rt^2=(r-\Gm)^2$ is the squared deviation from $\Gm$, in which $\Gm$ is the average value of $r$. This expression remains a three-parameter distribution because, as noted above, $\Gm=(a-\at)/2\Ga$. 

With this set of parameters, the affine-invariant scale is
\begin{equation}\label{eq:Tr2}
  \trr=e^r-\frac{\at}{\Gl}r+\frac{\Ga}{\Gl}\rt^2.
\end{equation}
Note that $\tr$ and $w$ are related by \Eq{wmetric}. We are using $w$ from \Eq{wr} and $\Gb=1$, as noted below \Eq{wr}. We ignore the extra $-1$ term in $\tr$ of \Eq{wmetric}, because the canonical form of probability distributions is invariant to adding a constant to $\tr$. The tilde parameters of the distribution in \Eq{gammaLNRu} are interchangeable with the nontilde parameters of the identical distribution in \Eq{gammaLNN}. The tilde expressions focus on the invariances that will help us to interpret ecological pattern. The nontilde expressions describe pattern in terms of the classic forms for the gamma and lognormal distributions. 

By the standard theory of maximum entropy, $q_r$ maximizes entropy on the incremental scale $\dd r$ subject to a constraint on the average value of the defining affine-invariant scale, $\angb{\tr}_r$. That constraint is the linear combination of three constraints: the average abundance on the process scale, $\angb{n=e^r}_r$, the average demographic process value, $\angb{r}$, and the variance in the demographic process values, $\angb{\rt^2}$.

By maximum entropy, all of the information in Hubbell's mechanistic process theory of neutrality and the matching gamma-lognormal pattern reduces to maximum randomness subject to these three constraints. 

However, it is very unlikely that we would have derived the correct form by maximum entropy without knowing the answer in advance. This limitation emphasizes that maximum entropy provides deep insight into process and pattern, but often we need an external theory to guide our choice among various possible maximum entropy formulations. 

Put another way, maximum entropy and process oriented theories, such as Hubbell's model, often work together synergistically to provide deeper insight than either approach alone.  

\subsection{Invariance, information and scale}

Before turning to invariance and the gamma-lognormal pattern of neutral theory, it is useful to consider some basic properties of invariance and information \autocite{tribus61thermostatics,cover91elements}. In particular, this subsection develops our claim that the affine-invariant scale provides the deepest insights into the relations between pattern and process.

We start by noting that, in the general expression for probability distributions
\begin{equation*}
  q_z=ke^{-\Gl\trz},
\end{equation*}
the affine-invariant scale, $\trz$, is equivalent to a common expression for the information content in a measurement, $z$, as 
\begin{equation*}
  I_z=-\log q_z.
\end{equation*}
This expression follows from assuming that: information depends on the probability, $q_z$, of observing the measured value and not on the value itself; rarely observed values provide more information than commonly observed values; and the information in two independent measurements is the sum of the information in each measurement.

\begin{figure*}[t]
\centering
\includegraphics[width=5.5in]{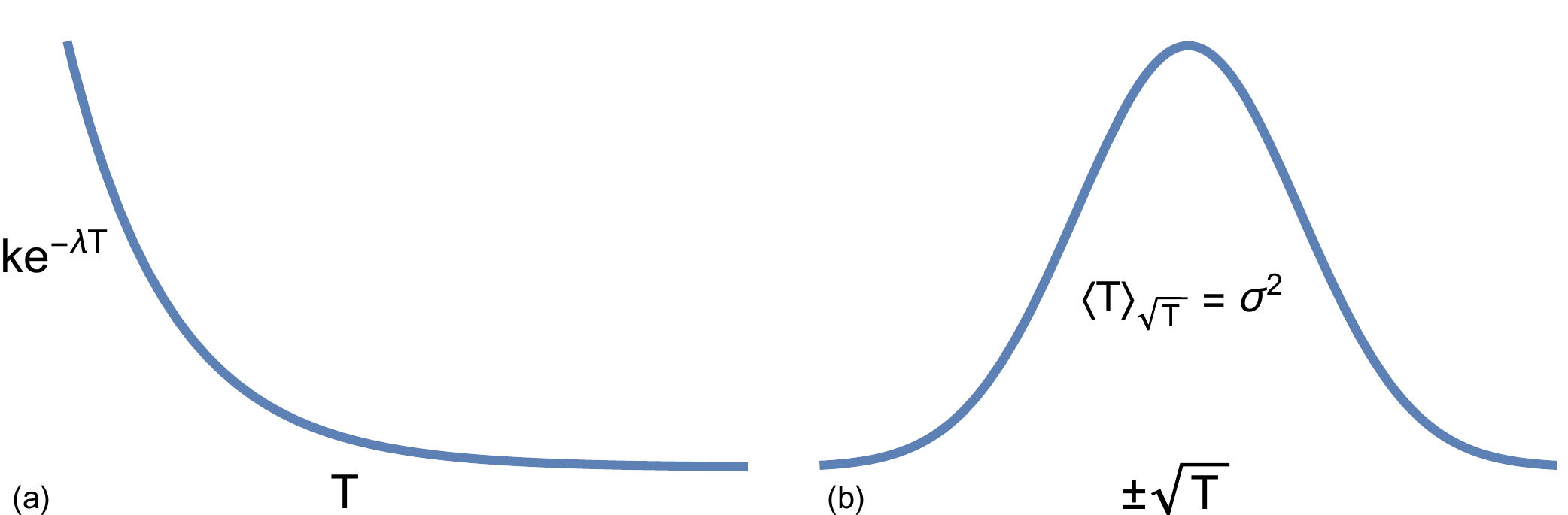}
\caption{Continuous probability distributions can often be expressed as exponential or normal distributions with respect to the affine-invariant scale. A continuous distribution typically can be written as $q_z=ke^{-\Gl\trz}$, from \Eq{Tz}. In the figure, $\tr\equiv\trz$. (a) A parametric plot of $q_z$ vs $\trz$ is exponential. All of differences between probability distributions are contained in the form of the affine-invariant scale, $\trz$. The change in information for each increment of the affine-invariant scale is $\Gl$, as in \Eq{constantInf}. (b) A parametric plot of $q_z$ vs $\pm\sqrt{\trz}$ is normally distributed when describing the deviations from a unimodal peak of $q_z$. The average of the deviations on the affine-invariant scale, $\langle\tr\rangle$, relative to measurements on the square root of that scale, $\sqrt{\trz}$, is the variance, $\Gs^2$. For the normal distribution, we can think of a deviation from the central location on the affine-invariant scale, $\trz=R_z^2$, as the squared radial deviations along the circumference of a circle with radius $R_z$, describing the squared vector length for an aggregation of variables. The variance is the average of the squared radial deviations relative to the scale of radial measures, $\sqrt{\trz}=R_z$. Most continuous unimodal distributions are, in this way, equivalent to a normal distribution when scaled with respect to the square root of the affine-invariant measure. See refs.~\citenum{frank16common,frank16the-invariances} for details.
}
\label{fig:canonical}
\end{figure*}

From the general expression for probability distributions
\begin{equation*}
  I_z=-\log q_z=-\log k+\Gl\trz.
\end{equation*}
Thus, an incremental change in information is equal to an incremental change in the affine-invariant scale
\begin{equation*}
  \dd I_z=-\dd\log q_z=\Gl\dd\trz.
\end{equation*}
Equivalently, the change in information with respect to a change in the affine-invariant scale,
\begin{equation}\label{eq:constantInf}
  \dovr{I_z}{\trz}=\Gl,
\end{equation}
is constant at all magnitudes of the measurement, $z$. Every measured increment on the $\trz$ scale provides the same amount of information about pattern. Constancy of information at all magnitudes is the ideal for a measurement scale. Thus, affine invariance provides the ideal scale on which to evaluate the pattern in measurements \autocite{frank14how-to-read}. \Figure{canonical} illustrates some key properties of the affine-invariant scale. 

Information is sometimes thought of as a primary concept. However, it is important to understand that, in this context, information and affine invariance are the same thing. Neither is intrinsically primary. 

We prefer to emphasize invariance, because it is an explicit description of the properties that pattern and process must obey \autocite{frank16the-invariances,frank16invariant,frank18measurement}. Further analysis of invariant properties leads to deeper insight. For example, only through invariance can we obtain the group theory expression for the canonical form of probability patterns (\Eq{canonical}).

By contrast, ``information'' is just a vague word that associates with underlying invariances. Further analysis of information requires unwinding the definitions to return to the basic invariances. 

\subsection{Invariance interpretation of the gamma-lognormal}

We turn now to the neutral theory model for abundances at local spatial scales. We showed that all of the information about pattern and process in the neutral theory is captured by the gamma-lognormal pattern in \Eq{gammaLNRu} as
\begin{equation*}
  q_r=ke^{-\Gl\trr}=ke^{-\Gl e^r + \at r - \Ga \rt^2},
\end{equation*}
which defines the affine-invariant scale in \Eq{Tr2} as
\begin{equation}\label{eq:LTr}
  \Gl\trr=\Gl e^r-\at r+\Ga\rt^2.
\end{equation}
On this scale, changes in $r$ provide the same amount of information about pattern at all magnitudes. Shifting the scale by a constant does not change the information about pattern in measurements. In other words, it does not matter where we set the zero point for $\trr$. Similarly, uniformly stretching or shrinking the scale, $\trr$, does not change the information in measurements of $r$. 

We can parse the terms of \Eq{LTr} with respect to constraint and invariance. When $r$ is large, the term $\Gl e^r=\Gl n$ dominates the shape of the distribution in the upper tail, which decays as
\begin{equation*}
  q_r\dd r=ke^{-\Gl e^r}\dd r=ke^{-\Gl n}\dd r
\end{equation*}
for sufficiently large $e^r=n$. The smaller the value of $\Gl$ relative to $\at$ and $\Ga$, the greater $e^r$ must be for this pattern to dominate. When $\Gl$ is relatively large compared with $\at$ and $\Ga$, this pattern dominates at all magnitudes and leads to the log series.

With respect to constraint, for large values of abundance, $n$, the constraint on average abundances dominates the way in which altered process influences pattern. With respect to invariance, a process that additively shifts or multiplicatively stretches the $e^r=n$ values does not alter the pattern in the upper tail. Similarly, pattern is invariant to a process that additively shifts process values, $r$, but processes that multiplicatively change $r$ alter pattern. Thus, we can evaluate the role of particular processes by considering how they change $n$ or $r$. 

The pattern at small and intermediate values of $r$ depends on the relative sizes of the parameters. If the $\at r$ term dominates, then the constraint, $\angb{r}$, on the average process value is most important. With respect to invariance when $\at r$ dominates, a process that additively shifts or multiplicatively stretches the $r$ values does not alter the pattern in the lower tail. That lower tail is a rising exponential shape, $e^{\at r}$, as in \Fig{neutralFit}c. 

When the $\Ga\rt^2$ term is negligible at all magnitudes, the combination of the dominance by $\at r$ in the lower tail, and the dominance by $\Gl e^r$ in the upper tail, yields the gamma distribution pattern on the abundance scale, $n$.

Finally, for magnitudes of $r$ at which the $\Ga\rt^2=\Ga\lr{r-\Gm}^2$ term dominates, the constraint, $\Gs^2=\angb{{r-\Gm}^2}$, on the variance in process values is most important. In this case, pattern follows a normal distribution, $e^{-\Ga\lr{r-\Gm}^2}$, on the $r$ scale, which is a lognormal distribution on the abundance scale, $n$. 

When combining numerous process values to obtain an overall net $r$ value, approximate rotational invariance is sufficient for the pattern to be very close to a perfect normal curve (see Introduction). When measuring net squared deviations from the mean, which is the squared radial distance, the pattern is invariant to shift and stretch of the squared radial measures, $\lr{r-\Gm}^2$. 

In practice, the lognormal pattern of abundance dominates when a constraint on $r$ dominates and net values of $r$ obey rotational invariance (symmetry) with respect to the summing up of the individual processes acting on abundance.

Any theory of process that leads to those three basic invariances will follow the gamma-lognormal pattern. The great unsolved puzzle is how specific mechanistic processes combine such that the structure of pattern is fully expressed by these particular invariances of pattern or, equivalently, by constraints on the average values of certain quantities in the context of maximum entropy. Our work opens the way for a more direct attack on this great puzzle by clarifying the anatomy of a pattern, thereby clarifying the puzzle that must be solved.

\section{The anatomy of pattern}

\begin{quote}
\em
[J]ust as the physiologist divides the animal world, according to anatomy, into families and classes, so the ornamentist is able to classify all pattern-work according to its structure [invariance]. Like the scientist, he is able even to show the affinity between groups to all appearance dissimilar; and, indeed, to point out how few are the varieties of skeleton upon which all this variety of effect is framed (ref.~\citenum{day87the-anatomy}, pp.~3--4). $\dots$ The fact of the matter is, the characteristic lines of time-honoured patterns are mainly the direct result of the restrictions under which the craftsman was working (ref.~\citenum{day87the-anatomy}, p.~47). 
\end{quote}

\noindent Invariances comprise the structural components in the anatomy of pattern. Commonly observed patterns almost always dissect completely into a few simple invariances. Our primary goal has been to introduce into ecological study the anatomy of pattern and the methods of dissection.

Identifying and naming the parts does not tell one how those parts came to be. In fact, common patterns are widespread exactly because so many different underlying mechanistic processes give rise to the same simple invariances.

Roughly speaking, one can think of a common pattern as an attractor. Each different underlying mechanistic process that develops into the generic form traces a distinctive path from some starting point to the generic endpoint of the attractor. All of the different mechanistic processes and starting points that end up at the same attractor form the basin of attraction for that pattern.

Our work characterized the anatomy of pattern---the anatomy of the attractors. The next step requires understanding how various combinations of mechanistic processes lead to one attractor or another. Equivalently, one can think of a mechanistic process as something that transforms inputs into outputs \autocite{frank13input-output}. Three questions follow. 

How do particular cascades of input-output transformations ultimately combine to produce overall transformations that associate with simple invariances? What separates some cascades from others with regard to association with different invariances? In other words, how can we assign different mechanistic cascades to one basin of attraction or another?

If we could answer those questions, then we could predict whether different mechanistic processes lead to the same pattern or to different patterns.

The fact that different processes can attract to the same pattern has been widely discussed in ecology \autocite{cohen68alternative,chave06comparing,adler07a-niche,zillo07the-impact,etienne07the-zero-sum,frank09the-common,chisholm10niche,mcgill10towards,matthews14neutral,matthews14fitting}. However, that past work typically did not explain common patterns in terms of invariance. Without invariance, one does not have a basis for describing the anatomy of common patterns or the reasons why certain processes attract to a particular pattern and others do not.

Invariance may provide a way to compare different models of process that lead to the same pattern. Among the many complex component processes that may occur in a model, which truly matter? In other words, which component processes shape the defining invariances and which are irrelevant? For the focal component processes of each model that matter, which empirical tests would tell us which of the alternative mechanistic models is the more likely match to natural processes? 

\section{Conclusions}

The apparent simplicity of invariance can mislead about its ultimate power. For example, probability patterns express a shift and stretch invariant scaling. That affine-invariant scaling provides a constant measure of information at all magnitudes. 

Shift and stretch invariance seem almost trivially simple. Yet, by analyzing how repeated transformations of shift and stretch retain invariance, we obtain the most general form that expresses various affine-invariant scales (\Eq{wmetric}). That affine symmetry group defines the simple, general structure of probability patterns and their uniform measurement scales.

Knowing the general invariant form of probability patterns reveals the relations between different approaches. Invariance provides powerful methods to analyze pattern and process.

To sum up, our invariance approach is not just another one among various alternatives. Rather, it is the only way to relate process to pattern, because the essence of pattern is invariance. Only by understanding what pattern actually is and how it generally arises can one begin to formulate testable hypotheses about mechanism. 

Put another way, pattern is always the interaction between, on the one hand, the generic aspects of invariance and scale that arise in all cases and, on the other hand, the particular aspects of biology that operate in each case. Without a clear view of that duality between the generic and the particular, it is easy to mistakenly attribute generic aspects of observed pattern to particular causes. To properly understand the role of specific mechanistic aspects in shaping pattern, one must evaluate pattern simultaneously from the perspectives of the generic and the particular.

\section*{Acknowledgments}

\noindent The Donald Bren Foundation (SAF) and the Swiss National Science Foundation (grant 31003A\_169671 to JB) support our research. SAF completed this work while on sabbatical in the Theoretical Biology group of the Institute for Integrative Biology at ETH Zürich.


\mybiblio	


\appendix
\renewcommand\thefigure{A\arabic{figure}}	
\setcounter{figure}{0}
\section{Appendix}

This appendix provides details of how to analyze changes of variable and changes of scale in a consistent way for discrete and continuous random variables. The material here was originally published at \href{https://doi.org/10.5281/ zenodo.2597895}{https://doi.org/10.5281/zenodo.2597895} under a \href{https://creativecommons.org/licenses/by/4.0/legalcode}{CC-BY} Creative Commons 4.0 license as a supplement to ref.~\citenum{frank19the-common}.

Discrete and continuous probability distributions are usually analyzed differently, which prevents a general understanding of scale. These notes present a technique by which a change of variable or a change of scale can be done in a consistent way for both discrete and continuous distributions \autocite{au99transforming}. The final section relates discrete and continuous scales, illustrated by the log series.

For example, suppose initial measurements are in terms of abundance, $n$, and we wish to analyze the data on the transformed logarithmic scale, $r=\log n$. How can we make the change of variable, $n\mapsto e^r$, consistently for discrete and continuous cases? 

The Dirac delta function provides the basis for a consistent method. The next section introduces the basic aspects and notation for the Dirac delta function. The following section shows how to use this method to obtain a consistent approach for transforming scale by change of variable. The final sections consider transformations between discrete and continuous variables and the specification of the domains of variables.

\subsection{Dirac delta function}

The Dirac delta function, $\Gd$, provides the key. The function is defined such that 
\begin{equation*}
  \int_{-\Ge}^{\Ge} \Gd(z)\dd z=1
\end{equation*}
for any real value $\Ge>0$. In other words, for any region of integration containing 0, the integral of $\Gd(z)$ is one. Then we also have
\begin{equation*}
  \int_{z-\Ge}^{z+\Ge} f(x)\Gd(x-z)\dd x=f(z).
\end{equation*}
In other words, the integral picks out the function evaluated at the point $x=z$, at which $\Gd(x-z)=\Gd(0)$.  

With that definition, we can write a discrete probability distribution at the set of points $\Omega=\left\{x_i\right\}$ as a continuous probability density function
\begin{equation}\label{eq:discrete1}
  f(x)\sum_{x_i\in\Omega}\Gd(x-x_i)=f(x)\Gd_x,
\end{equation}
because the cumulative distribution function, $F(x)$, of the continuous density, $f(x)\Gd_x$, has the form of a discrete probability distribution
\begin{equation*}
  F(a)=\int_{-\infty}^{a+\Ge} f(x)\Gd_x\dd x=\sum_{x_i\in\Omega} f(x_i),
\end{equation*}
in which $x_i<a+\Ge$ for an infinitesimal positive value, $\Ge$. We need the extra $\Ge$ so that $\Gd_x$ integrates to one around a point $x_i=a$, 

For continuous distributions, let the density of points in $\Omega$ increase to fill the interval $(-\infty,\infty)$ continuously. Then
\begin{equation*}
  \Gd_x=\sum_{x_i\in\Omega}\Gd(x-x_i)\rightarrow\int_{-\infty}^\infty \Gd(x-x_i)\dd x_i=1,
\end{equation*}
because, whatever the value of $x$, there will be some point $x_i\in\Omega$ for which $x=x_i$, and any integral over the region including that point is one. With $\Gd_x=1$, the continuous probability density function is $f(x)$, and the cumulative distribution function is
\begin{equation*}
  F(a)=\int_{-\infty}^{a+\Ge} f(x)\dd x=\int_{-\infty}^{a} f(x)\dd x,
\end{equation*}
because $\int_{a}^{a+\Ge} f(x)\dd x=0$ for infinitesimal $\Ge$ and finite $f(x)$. 

\subsection{Change of variable}

We seek a consistent method for doing a change of variable in both continuous and discrete probability distributions. The approach arises from always considering the probability associated with a value as the area of a rectangle. 

For a continuous distributions, we write $f(x)\dd x$, which is the product of the probability function, $f$, as the height, and the infinitesimal interval, $\dd x$, as the width. Thus, the probability in the interval $(a,a+\Ge)$, with small width $\Ge$, is
\begin{equation*}
  \int^{a+\Ge}_a f(x)\dd x\approx f(a)\Ge,
\end{equation*}
the product of the height, $f(a)$, and the width, $\Ge$. When we change variables, $x\mapsto g(x)\equiv y$, we obtain both a new height, $f(y)$, and a new width, $\dd y$, and so we must compensate appropriately, as shown below.

For discrete distributions, we may write $f(x)\GD x$, in which $\GD x$ is the width associated with a discrete point, $x$. Typically, we assume that $\GD x=1$ for all $x$, and write the discrete probability as $f(x)$. We can think of this as the area of a rectangle with implicit width of one. 

When we change variables, $x\mapsto g(x)\equiv y$, traditionally one keeps the interval widths, $\GD y=1$, as one on the new scale, $y$, and the probabilities are simply $f(y)$ at the new points, $y$. However, by changing scales, the spacing between the discrete points on the $y$ scale differs from the spacing between points on the original $x$ scale. 

This change of spacing can be interpreted as a change in the widths associated with discrete probability points, or as a change in the density of probability points in intervals along the $y$ scale. Thus, as in the continuous case, we may wish to keep track of how both the heights change, $f(x)\mapsto f(y)$, and how the widths change with a change of scale, $\GD x\mapsto \GD y$. Doing so provides a consistent way of changing variables for continuous and discrete cases.

We begin with the standard method for continuous variables. We then demonstrate an analogous method for discrete variables based on the Dirac delta function.

For a continuous distribution, $f(x)$, we make the change $x\mapsto g(x)\equiv y$. With that transformation, we have
\begin{equation*}
  \dd y/\dd x=g'(x).
\end{equation*}
Define
\begin{equation*}
  m_y = \frac{1}{\abs{g'(x)}},
\end{equation*}
in which the absolute value arises because we are using $\dd x$ and $\dd y$ as positive probability measures. Thus,
\begin{equation*}
  \dd x = m_y\dd y.
\end{equation*}
Then the standard result for the change of variable $x\mapsto y$ in a continuous distribution yields
\begin{equation}\label{eq:contChange}
  f(x)\dd x = f(y)m_y\dd y.
\end{equation}

For discrete distributions, we will derive the analogous change of variable expression
\begin{equation}\label{eq:discreteChange}
  f(x)\Gd_x\dd x = f(y)m_y\Gd_y'\dd y=f(y)\Gd_y\dd y.
\end{equation}
To obtain this result, we need to show that the change of variable $x\mapsto g(x)\equiv y$ leads to
\begin{equation*}
  \Gd_x\dd x\mapsto m_y\Gd_y'\dd y=\Gd_y\dd y,
\end{equation*}
which follows if 
\begin{equation*}
  \Gd_x\mapsto m_y^{-1}\Gd_y\equiv\Gd_y'.
\end{equation*}
To obtain this expression for $\Gd_y'$, we need the general change of variable rule for the Dirac delta function
\begin{align*}
  \Gd(x-x_i)&\mapsto\abs{g'(x_i)}\Gd\lrb{g(x)-g(x_i)}\\[4pt]
  			&=m_y^{-1}\Gd(y-y_i).
\end{align*}
Thus, with $\Omega'=\left\{g(x_i)\right\}=\left\{y_i\right\}$, we have
\begin{equation*}
  \Gd_x=\sum_{x_i\in\Omega}\Gd(x-x_i)\mapsto
  	m_y^{-1}\sum_{y_i\in\Omega'}\Gd(y-y_i)=m_y^{-1}\Gd_y.
\end{equation*}

\subsection{The gamma and log series distributions}

The gamma distribution is given by the probability function
\begin{equation*}
  f(x)=kx^{\Ga-1}e^{-\Gl x},
\end{equation*}
in which the constant $k$ normalizes the total probability to be one. With $\Ga=0$ and $x>x_0>0$ for $x_0$ not too close to zero, this has the same mathematical form as the log series distribution. 

Consider the change in variable $r=\log x=g(x)$, which corresponds to $x\mapsto e^r$. Then,
\begin{equation*}
  \abs{g'(x)}=\dd\log x/\dd x=1/x=e^{-r}=m_r^{-1}.
\end{equation*}
If we consider $f(x)$ as a continuous distribution, then we can apply the formula for change of variable in \Eq{contChange} to obtain
\begin{align*}
  f(x)\dd x &= f(r)m_r\dd r\\[4pt]
  			&=ke^{r(\Ga-1)}e^{-\Gl e^r}e^r\dd r\\[4pt]
        	&=ke^{r\Ga-\Gl e^r}\dd r\\[4pt]
        	&=h(r)\dd r,
\end{align*}
in which
\begin{equation}\label{eq:hr}
  h(r)=f(r)m_r=ke^{r\Ga-\Gl e^r},
\end{equation}
for $r>\log x_0$. For $\Ga=0$, transforming the log series form 
\begin{equation*}
  f(x)=kx^{-1}e^{-\Gl x}
\end{equation*}
by $r=\log x$ yields the equivalent distribution on the $r$ scale as
\begin{equation}\label{eq:hrlogseries}
  h(r) =ke^{-\Gl e^r}.
\end{equation}
Now consider $f(x)$ as a discrete distribution. Then, by \Eq{discreteChange}, we immediately have
\begin{equation*}
   f(x)\Gd_x\dd x = f(r)m_r\Gd_r'\dd r=h(r)\Gd_r'\dd r,
\end{equation*}
in which the full form of $h(r)$ is given in \Eq{hr}, and we also have the log series form with $\Ga=0$ in \Eq{hrlogseries}. 

Thus, by using the measure $\dd r$ for the widths in the continuous case and the measure $\Gd_r'\dd r$ for the widths in the discrete case, we obtain the identical probability function $h(r)$ for the continuous and discrete cases. 

For the discrete case, we must keep in mind that
\begin{equation*}
  h(r)\Gd_r'\dd r=h(r)m_r^{-1}\Gd_r\dd r,
\end{equation*}
in which the right side is the traditional expression that picks out the probability mass, $h(r)m_r^{-1}$, as the heights at the points defined by $\Gd_r\dd r$, implicitly using constant widths of one on all scales, because at a discrete point, $r^*$, at which the probability is nonzero, 
\begin{equation*}
  \int_{r^*-\Ge}^{r^*+\Ge}\Gd_r\dd r=1.
\end{equation*}

In the gamma example, $m_r^{-1}=e^{-r}$, with $x=1,2,\dots$ and $r=\log 1, \log 2, \dots$, we have the traditional expression for a discrete change of variable with constant widths of one as
\begin{align*}
  f(x)\Gd_x\dd x&\mapsto h(r)m_r^{-1}\Gd_r\dd r\\[5pt]
  	&=ke^{r(\Ga-1)-\Gl e^r}\Gd_r\dd r.
\end{align*}
For the log series case, $\Ga=0$, this becomes
\begin{equation*}
  h(r)m_r^{-1}\Gd_r\dd r=ke^{-r-\Gl e^r}\Gd_r\dd r.
\end{equation*}
We can go back to the classic log series expression by reversing the change, $e^r\mapsto n$, yielding the discrete distribution $f(n)\Gd_n\dd n$ for $n=1,2,\dots$, with
\begin{equation*}
  f(n)=k\nmo e^{-\Gl n}.
\end{equation*}

In these examples, we have assumed that the distributions on the $r$ and $n\equiv x$ scales are either both discrete or both continuous. In application, it will usually make sense to think of process, $r$, as a continuous variable, and abundance, $n$, as a discrete variable. Therefore, we need to consider transformations between continuous and discrete variables. We discuss that topic in the final section, after a brief summary of the discrete transformations.

\subsection{Summary of alternative discrete expressions}

We have two different ways of expressing transformed discrete variables, in which the initial distribution is given by $f(x)\Gd_x\dd x$, and we transform $x\mapsto y$. 

In the first expression, the transformed distribution is 
\begin{equation*}
  f(y)m_y\Gd_y'\dd y=h(y)\Gd_y'\dd y,
\end{equation*}
in which $h(y)=f(y)m_y$ is the same expression as obtained when transforming continuous variables. The measure for the $y$ scale, $\Gd_y'\dd y$, stretches or shrinks in relation to the measure for the $x$ scale, altering the widths associated with each height. This form has the advantage of retaining the same expressions for the probability functions in the discrete and continuous cases.

In the second expression, the transformed distribution is 
\begin{equation*}
  f(y)\Gd_y\dd y=h(y)m_y^{-1}\Gd_y\dd y,
\end{equation*}
in which this expression highlights the difference between the standard form of the probability function obtained in the discrete case, $f(y) = h(y)m_y^{-1}$, and the standard form of the probability function obtained in the continuous case, $h(y)=f(y)m_y$. Here, we assume the widths associated with each probability point remain one on all scales. 

For the transformation $x\mapsto g(x)\equiv y$, the value $m_y^{-1}=\abs{g'(x)}$ determines the distinction between the discrete and continuous cases, associated with the change in widths between scales.

\begin{figure*}
\centering
\includegraphics[width=\textwidth]{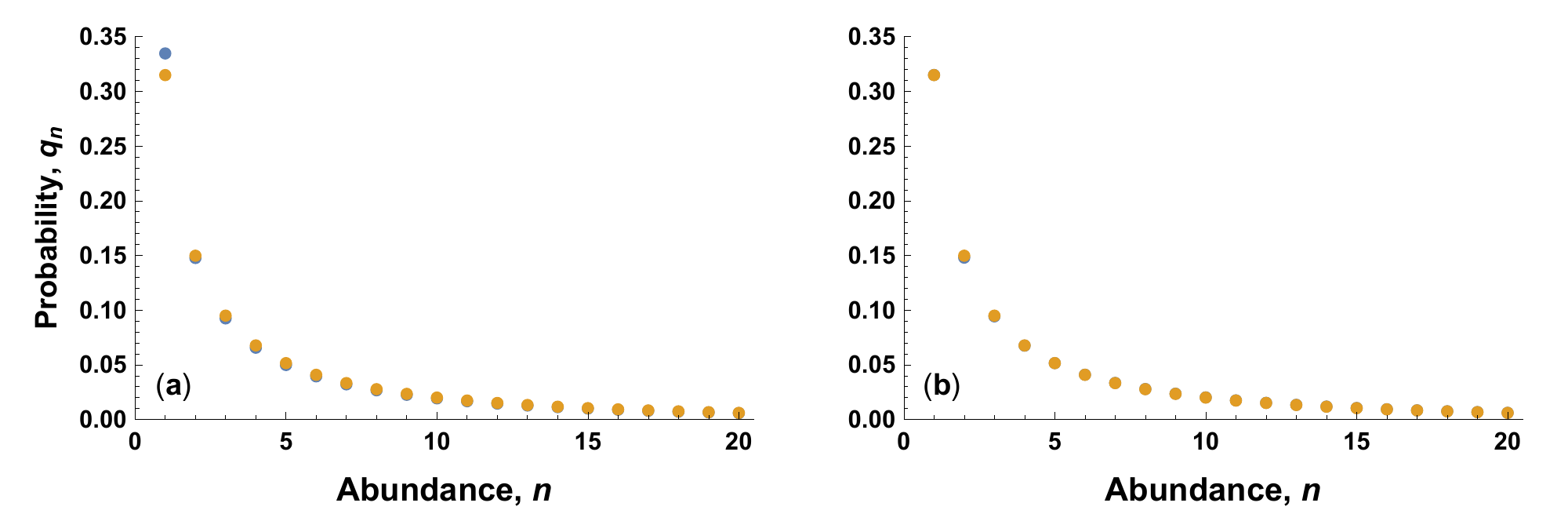}
\caption{Close match between the continuous distribution for process, $r$, and the discrete log series for abundance, $n$. The blue circles show the probabilities of the discrete values of $n$ obtained from the continuous distribution in $r$, calculated by \Eq{cont2disc}. The gold circles show the actual values of $q_n$ for the log series in \Eq{apxqn}. For most points, the values are nearly identical, causing the gold circles to hide the underlying blue circles. (a) When using no offset for continuous intervals, $\Gg=0$, a slight mismatch occurs, particularly at $n=1$. The nonlinearity of $q_r$ causes the mismatch. (b) When using an offset of $\Gg=0.1$, the continuous distribution of the process, $r$, maps almost perfectly onto the discrete log series of abundance, $n$. For all calculations, $\Gl=0.05$.}
\label{fig:cont2discr}
\end{figure*}

\subsection{Transforming between continuous and discrete variables}

In the prior cases, we transformed from one discrete variable to another discrete variable or from one continuous variable to another continuous variable. The expressions for transformation followed without any further assumptions.

In the case of the log series and the distribution of abundances, it often make sense to consider process, $r$, as a continuous variable, and abundance, $n$, as a discrete variable. We usually think of process as causing abundance. So we should begin with the continuous distribution for process, $q_r$, and seek the corresponding discrete distribution for abundance, $q_n$.

The probability mass, $q_n$, at a particular value of $n=1,2,\dots$, should map to the total probability for a matching range of growth rates, such that
\begin{equation}\label{eq:cont2disc}
  q_n=\int_{a_n}^{b_n} q_r\dd r,
\end{equation}
in which $r>a_1$. 

We need the particular form of $q_r$, which we take as the fundamental shift-invariant distribution in the main text
\begin{equation}\label{eq:apxqr}
  q_r=ke^{-\Gl e^r}.
\end{equation}
We also need, for each $n$, the interval of growth rates, $(a_n,b_n)$, that maps to the abundance, $n$. In particular, we need a sequence of contiguous intervals, $\left\{\lr{a_n,b_n}\right\}$, that associate each value of $n$ to an interval for $r$, such that $a_{n+1}=b_n$. 

The problem concerns how to pick the sequence of intervals. The simplest approach is to use a standard rounding procedure, such that 
\begin{align*}
  a_n&=\log(n-0.5+\Gg)\\
  b_n&=\log(n+0.5+\Gg),
\end{align*}
in which $\Gg<0.5$ is a correction for centering intervals to account for the nonlinearity in the mapping between the continuous and discrete probability expressions.

If we use $\Gg=0.1$, the transformation from the continuous scale $r$ to the discrete scale $n$ in \Eq{cont2disc} yields a distribution that closely matches the log series
\begin{equation}\label{eq:apxqn}
   q_n\approx kn^{-1}e^{-\Gl n}.
\end{equation}
\Figure{cont2discr} shows the match between the continuous distribution for $r$ and the associated discrete distribution for $n$.

\end{document}